**ENERGY CONSUMPTION | ENERGY**

Avoiding combustion significantly reduces primary energy consumption

# Electricity instead of heat

Axel Kleidon | Harald Lesch

*The energy transition is also about switching to electricity-based technologies such as heat pumps and electric mobility. They avoid heat as an intermediate step and are therefore much more efficient. This can significantly reduce the demand for primary energy in the future, which can then be fully covered by the expansion of renewable energies. Entropy and the maximum possible combustion temperature can be used to understand why combustion is so inefficient.*

The need for an energy transition towards renewable energy is typically motivated by global climate change, as it is mainly caused by burning fossil fuels. These still account for around 80 % of Germany's primary energy consumption [1], 70 % of which is imported at an annual cost of around 80 billion euros [2]. This represents a huge sum - about 20 % of the federal budget. In order to replace fossil fuels, counteract climate change and become less dependent on huge energy imports, we should replace them with sustainable and climate-neutral energy sources.

The fact that the energy transition is focussing on more modern technologies that use electricity and avoid combustion and heat as an intermediate step, which increases their efficiency enormously, is being pushed into the background. Efficiency is defined here like the classic efficiency of a heat engine: Utilisation is more efficient the less energy has to be expended to achieve the same goal, such as the generation of light, motion or space heating. The maximum possible efficiency of heat engines is determined by the second law of thermodynamics, i.e. entropy. This also leads us to the question of how these increases in efficiency can be explained in physical terms with the help of entropy.

## Increased efficiency in light generation

The technical history of light generation (Figure 1) gives us an initial impression of such an increase in efficiency. Originally, combustion was used for this purpose. The chemical energy of wax is burnt in a candle, i.e. converted into heat. The rising soot particles are so hot that they radiate some of their heat in the visible range, thus generating light. The efficiency is low, however, as the flame temperature is relatively low at around 1400 °C. The heat released is mainly emitted in the infrared, i.e. radiation that does not reach the target - visible light.

The next step in development was the light bulb. It uses electricity as an energy source and produces light by heating the filament, which then lights up. As the temperature of up to 3000 °C is higher than that of a candle flame, the emission is shifted further into the visible range. Incandescent lamps are therefore more efficient than candles, even if the majority of the energy is

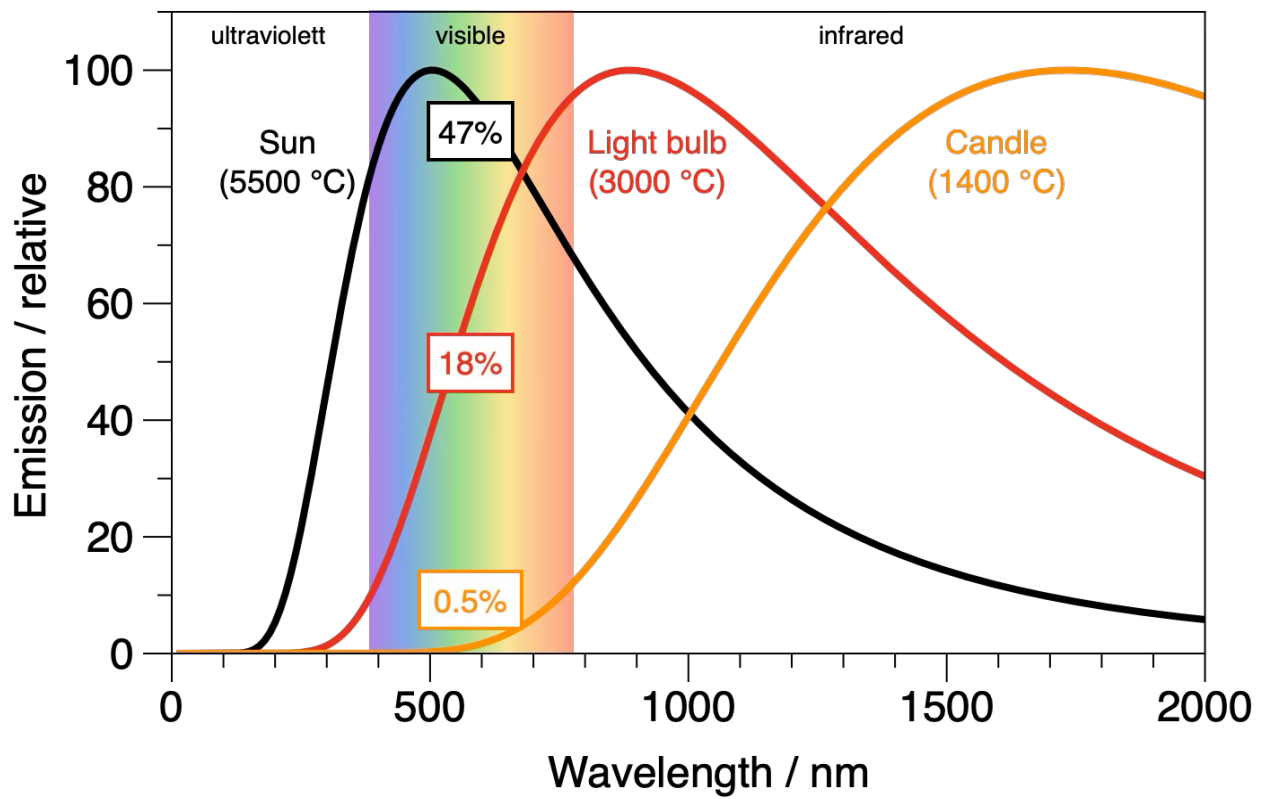

**FIGURE 1: LIGHT GENERATION**

*Candles and light bulbs generate light via heat and thermal emission. As a large proportion of this occurs in the infrared wavelength range, generating light via heat is associated with unavoidable losses - even the sun emits a significant proportion in the non-visible spectrum. In contrast, LEDs emit directly visible light, avoid heat as an intermediate step and are much more efficient.*

still emitted in the infrared. Even at even hotter temperatures, for example the sun with its surface temperature of 5500 °C, a substantial part of the emission occurs in the infrared, i.e. non-visible range. Light generation via thermal radiation is obviously associated with unavoidable inefficiency.

Nowadays, light generation is dominated by LEDs. Although electricity is also used as an energy source here, light is generated by electroluminescence, a quantum effect, instead of heat. By selecting the appropriate band gap in the semiconductor, light is only emitted in the desired wavelengths. Heat is essentially generated solely by the electrical resistance of the semiconductor material through which the current flows. LEDs are therefore fundamentally much more efficient because heat is avoided as an intermediate step in the light generation process.

The history of the candle, the light bulb and the LED exemplifies the clear development towards greater efficiency, because more light is produced for the amount of energy used. Combustion and heat as intermediate steps are avoided, the development has moved towards the utilisation of electricity and quantum physical effects. We can expect a similar development in other technologies associated with our energy requirements, namely that combustion will also be replaced in the future for electricity generation, room heating, mobility and other processes.

## Why combustion is inefficient

But why is combustion, i.e. the intermediate step via heat, so inefficient? In a power plant, we can recognise this from the well-known Carnot efficiency (see "The second law and energy conversions with heat"). The waste heat - i.e. the unconverted energy - enables entropy to be exported so that the energy conversion can follow the second law of thermodynamics. The greater the temperature difference between combustion and waste heat, the greater the maximum possible efficiency and the more of the released heat can theoretically be utilised. A power plant therefore becomes more efficient when utilising a greater temperature difference.

Temperature plays a key role in determining how efficient a combustion process is, for example if it is to perform mechanical work. The maximum possible temperature in a combustion process is described by the adiabatic flame temperature, at which the process is in principle reversible. The heat that is then released during the reaction has the lowest thermal entropy. This follows directly from Clausius' expression, according to which the entropy change $\Delta S$ of a system depends on the heat supplied $\Delta Q$ and the prevailing temperature $T$: $\Delta S = \Delta Q/T$. However, as we cannot utilise such high temperatures in many cases, we waste a considerable amount of the fuel's low entropy, which could in principle be used to perform work.

The maximum combustion temperature can be estimated as follows. We assume a certain chemical reaction in which a certain amount of heat $\Delta Q$ is released. Combustion requires a certain amount of reactants, which is determined by the stoichiometric ratio of the reaction. These reactants enter into the reaction at room temperature and the products are heated by the released heat. The resulting temperature can be determined from the heat capacity of the substances involved and the heat released. If we assume the minimum quantity of substances in an ideal combination that is required for combustion, we can then determine the maximum temperature of the products, i.e. the maximum combustion temperature.

Let's take the combustion of methane as an example. For one mole of methane ($CH_4$), two moles of oxygen ($O_2$) are needed, 890 kJ of heat are released per mole, and one mole of carbon dioxide ($CO_2$) and two moles of water ($H_2O$) are produced. In general, however, combustion takes place with air, where each mole of oxygen is accompanied by around four moles of nitrogen. Although the latter are chemically uninvolved, they contribute to the heat capacity. This gives us the mass equation

$$CH_4 + 2O_2 + 8N_2 \rightarrow CO_2 + 2H_2O + 8N_2. \tag{1}$$

To determine the maximum combustion temperature, we need the heat capacity of the gases involved. The molar heat capacity for diatomic gases is around 30 J mol$^{-1}$ K$^{-1}$, which we use here as an approximation. So if we consider the combustion of 1 mol methane, a total of 11 mol are involved, so the heat capacity is 330 J K$^{-1}$. If the reactants enter combustion at a room temperature of 20 °C, we obtain the maximum combustion temperature via the thermal energy balance per mole of methane:

$$330 \text{ J K}^{-1} \cdot 293 \text{ K} + 890 \text{ kJ} = 330 \text{ J K}^{-1} \cdot T_{max}. \tag{2}$$

This yields a temperature of $T_{max} \approx 3000$ K or 2700 °C. This calculation is only an approximation because the chemical equilibrium shifts at high temperatures and therefore complete combustion does not occur. An even higher temperature could be achieved when burning with pure oxygen,

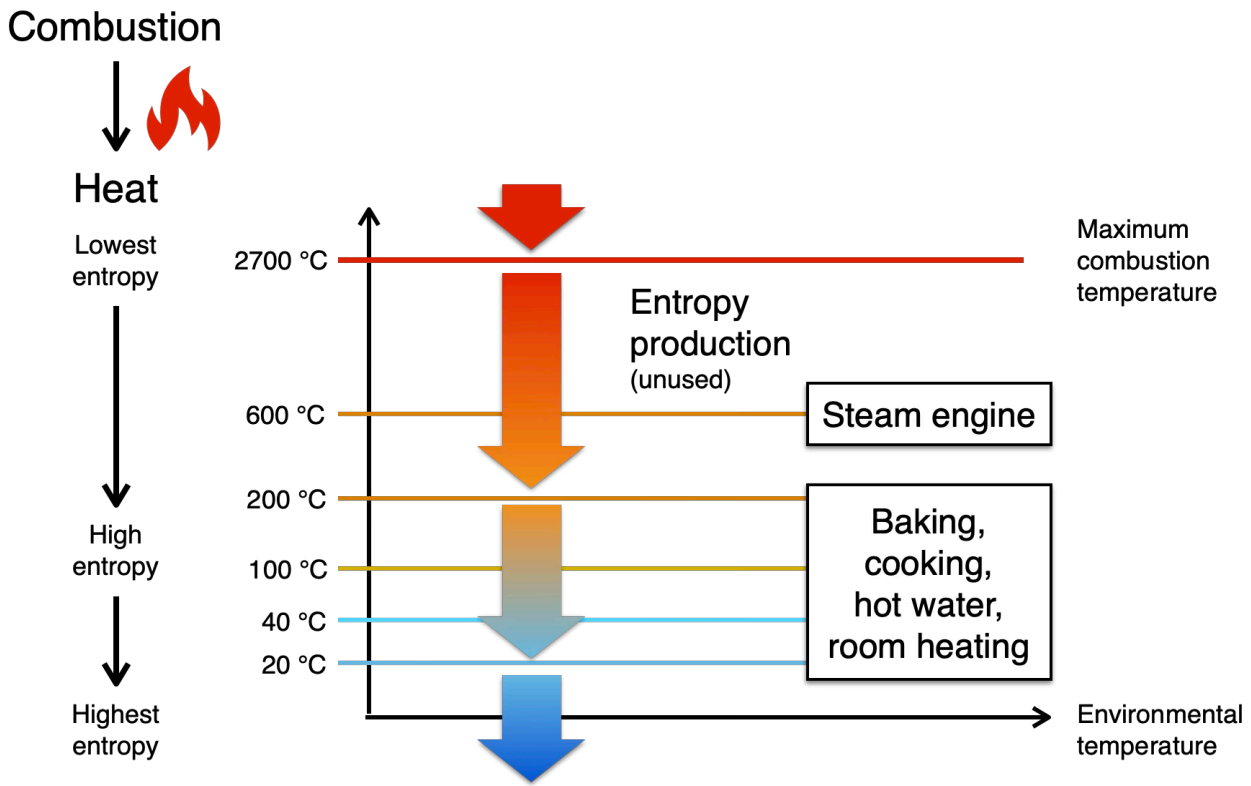

**FIGURE 2: LOSSES DURING COMBUSTION**

*Increase in entropy starting from the maximum temperature (above, example of methane as a fuel) that can be reached during combustion, towards the temperatures at which heat is utilised. Since the high temperature during combustion is generally not utilised, especially in households, a lot of entropy is generated during combustion. As a result, combustion-based technologies are typically inefficient.*

as the heat capacity is lower and the combustion temperature can then be correspondingly higher.

This high combustion temperature is so relevant because the heat released in the process has very low entropy, so in principle a lot of usable energy can be recovered from it. As soon as combustion is no longer ideal, i.e. more air is supplied for combustion than necessary, the heat is distributed over more mass, which leads to a lower temperature and higher entropy. With each further step that brings the heat to lower temperatures, the entropy increases accordingly - it is increased by the mixing of the hot heat with the cool environment.

Although the energy is retained, it becomes less usable, meaning it can do less work (Figure 2). As the high combustion temperatures are generally not achieved or needed - around 20 °C is the target for generating room heating - combustion is therefore associated with a high degree of inefficiency. Combustion processes cannot effectively utilise the ability of the high-quality energy in the fuel to perform work.

The inefficiency of combustion becomes even more obvious when we consider that the hydrocarbons we currently burn were produced by photosynthesis during Earth's history. Photosynthesis has a very low efficiency of about 1%, converting sunlight into chemical energy in the form of biomass [3]. Together with the inefficiency of combustion, this classic type of energy

## The second law and energy conversions with heat

### a. Power plant

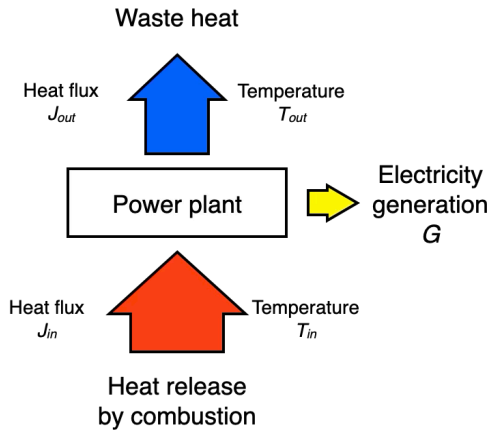

### b. Heat pump

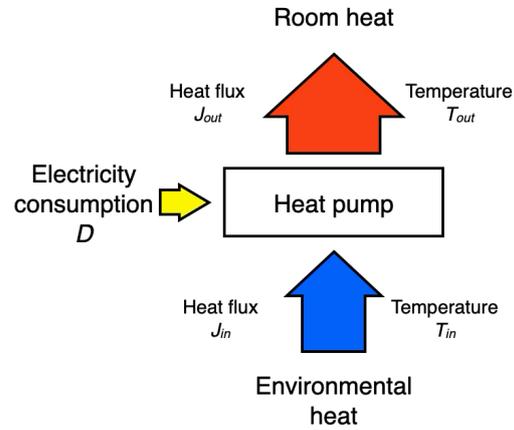

Thermal power stations and combustion engines generate usable energy such as electricity or motion from heat. The reverse is also true, as is the case with refrigerators, air conditioning systems and heat pumps: they utilise electricity to convert it into heat or cold. These technologies follow the laws of thermodynamics. Thermodynamics limits the maximum amount that can be converted. We show this below for both directions.

### Power plant: electricity from heat

In a power plant, heat is released by combustion at a certain rate $J_{in}$, from which usable energy $G$ is generated and waste heat $J_{out}$ is produced. The first law requires the conservation of energy during conversion, i.e.

$$J_{in} = J_{out} + G. \tag{1}$$

At the same time, the second law must also be fulfilled. This requires that at least as much entropy is exported via waste heat as is supplied via combustion. We can express this condition using Clausius' formula for entropy changes, $\Delta S = \Delta Q/T$, as

$$J_{out}/T_{out} \geq J_{in}/T_{in}, \tag{2}$$

where $T_{in}$ is the combustion temperature and $T_{out}$ is the temperature of the waste heat. The usable energy $G$ does not appear in this entropy balance, as it is not associated with entropy - it is therefore free for any utilisation. Ideally, both terms in (2) are equal, in which case there is no loss in the form of entropy production during the conversion process. We can then solve (2) for $J_{out}$, insert it into (1) and solve for the maximum power $G$. This gives us the well-known expression for the power limit of a heat engine:

$$G_{max} = J_{in} \cdot (T_{in} - T_{out})/T_{in}. \tag{3}$$

The efficiency of electricity generation is given by the Carnot efficiency, $(T_{in} - T_{out})/T_{in}$. From the expression we can recognise: The higher $T_{in}$ and the greater the temperature difference ($T_{in} - T_{out}$), the more heat can be converted into usable energy.

### Heat pump: heat from electricity

A heat pump is practically the opposite of a thermal power station. It consumes usable energy at a rate $D$ and absorbs heat at a rate $J_{in}$ from the environment at a temperature $T_{in}$ - like the waste heat from a power plant. It thus heats the room air at a rate $J_{out}$ at temperature $T_{out}$ - like combustion in a power plant. Here too, the laws of thermodynamics limit the maximum amount of heat a heat pump can generate from a certain power consumption. Again, the first law applies:

$$D + J_{in} = J_{out}, \tag{4}$$

and the second law:

$$J_{out}/T_{out} \geq J_{(in)}/T_{in}. \tag{5}$$

Again, we can determine the ideal case. To do this, we solve (5) for $J_{in}$ and insert this into (4). This gives us the maximum heat generation that is possible with a certain consumption rate $D$:

$$J_{out,max} = D \cdot T_{out}/(T_{out} - T_{in}). \tag{6}$$

We can see that heat generation is proportional to power consumption, but is amplified by the factor $T_{out}/(T_{out} - T_{in})$. This factor is also known as the *coefficient of performance.* It is always greater than one, i.e. greater than heat generation through simple dissipation, such as in an electric water kettle. While real coefficients of performance of around 3-5 are far below the theoretical limit, we can recognise from the expression: The smaller the temperature difference, the less electricity is needed to generate a certain amount of heat.

conversion is therefore clearly inferior to photovoltaics with its 20 % efficiency, which is already standard today.

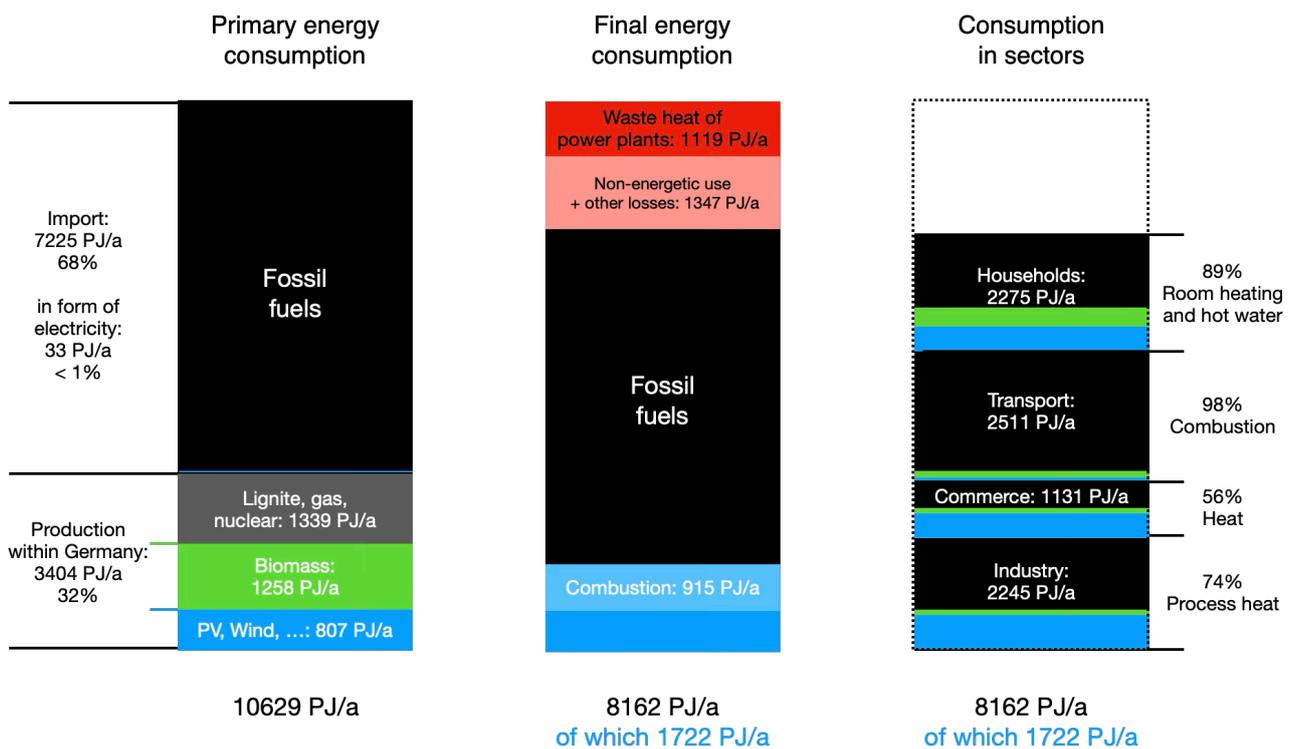

*FIGURE 3: PRIMARY ENERGY CONSUMPTION IN GERMANY*
*Role of the combustion of fossil fuels in Germany's current energy consumption. The diagram shows the share of fossil (black) and renewable (green) fuels as well as electricity (blue) in primary energy consumption (left), in final energy consumption (centre), i.e. after taking into account electricity generation in power plants and their waste heat (red) and other losses, as well as the breakdown of consumption into four sectors (right) (data source: AGEB [1]).*

## Combustion currently dominates

Our energy system in Germany is currently heavily based on combustion, even if renewables have already significantly displaced thermal power plants in electricity generation. However, mobility still mainly consists of combustion engines, we heat rooms and water predominantly by means of combustion, and industry uses combustion to obtain process heat for the production of various products. We can see from the composition of Germany's primary energy consumption (Figure 3) just how prominent a role combustion still plays at present.

The starting point of the diagram is primary energy consumption (Figure 3, left). The import of fossil fuels, i.e. coal, oil and gas, dominates consumption and therefore combustion. They account for around 80 % of the primary energy consumption of Germany, 68 % of which is imported [1]. The smaller share is contributed by domestic production from lignite (around 9 %) and renewables. However, we must bear in mind that both the fuels and the directly generated renewable electricity are added together - in other words, the energy before conversion and in an already usable form are added together. Of the renewables, only biomass or biogas (green) provides a fuel that is used for combustion. The other forms, such as photovoltaics, wind and water, generate electricity directly (blue), i.e. usable energy. This difference is indicated by the different colours in Figure 3.

In conventional electricity generation in thermal power plants, the losses due to conversion are directly recognisable in the illustration. Here, it is also entropy that explains these losses and why they are unavoidable (see "The second law and energy conversions with heat"). On the one hand, the maximum possible combustion temperature is generally not reached, and on the other hand, the well-known maximum efficiency of a heat engine comes into play. Both of these factors mean that considerably less than the heat of combustion is converted into electrical energy. Current thermal power stations typically have efficiencies of around 40 %, with the rest being released into the environment as waste heat via the cooling towers. Modern gas and steam combined cycle power plants can achieve up to 60% because they can better utilise the combustion temperature in combination with a steam turbine.

The other areas of our energy system are summarised in this diagram in final energy consumption in the form of four sectors (Figure 3, right): Households, transport, industry, and commerce, trade and services. However, this diagram does not show the further conversions within the sectors. Combustion and heating also play a dominant role here. Households use around 89 % of energy for room heating and hot water, while the figure for commerce, trade and services is 56 %. In the transport sector, almost all energy is used for heat generation by combustion engines. This typically generates more than 80 % waste heat and less than 20 % is converted into motion. In industry, 74 % of energy is used for process heat in the production of steel, glass, paper or chemical raw materials. Here, heat is required at high temperatures, for example to melt raw materials or to shift the chemical equilibrium in reactions. Efficiency gains are less easy here because high temperatures are required.

Let's summarise: 92% of our primary energy consumption is characterised by fuels and combustion. However, if we generate electricity directly, as is the case with photovoltaics, wind energy and hydropower (the remaining 8%), we can avoid the huge conversion losses that inevitably occur during combustion. So if we switch to electricity-based technologies, our electricity demand will increase, but our energy system will be more efficient and primary energy consumption will be lower.

## More efficient without combustion

We have already seen the higher efficiency achieved by avoiding combustion and heat in the generation of light. This increase in efficiency is also very clear in electricity generation (Figure 4). At the beginning of the 1990s, 94.7 % of electricity was generated by thermal power plants [1, 4], i.e. by combustion or nuclear energy. However, only 36.3 % of the primary energy used was converted into electricity, the rest was waste heat. This picture has changed dramatically with the expansion of photovoltaics and wind. Thermal power plants only contributed 52.6 % to electricity generation in 2023, with the rest being generated by solar, wind and hydropower without heat and waste heat. This 52.6 % for thermal power plants also includes electricity generation from the combustion of biomass, which contributes 9.7 % as a renewable form, while the remaining 42.9 % is accounted for by fossil fuels (Figure 4 above).

Overall, electricity generation in Germany has become much more efficient due to the displacement of heat as an intermediate step: The ratio of electricity generation to primary energy used for this purpose has risen from 36.3 % in 1991 to 58.3 % in 2023 (Figure 4 below). Specifically, the amount of primary energy required to generate electricity has fallen from 5413 PJ per year in 1991 to 3107 PJ per year in 2023. The generation of waste heat has fallen accordingly, even though electricity generation has not changed significantly. This is exactly what we would expect has happened - our electricity generation has become more efficient because combustion has been displaced. Since 1991, a good 40 % of primary energy has been saved!

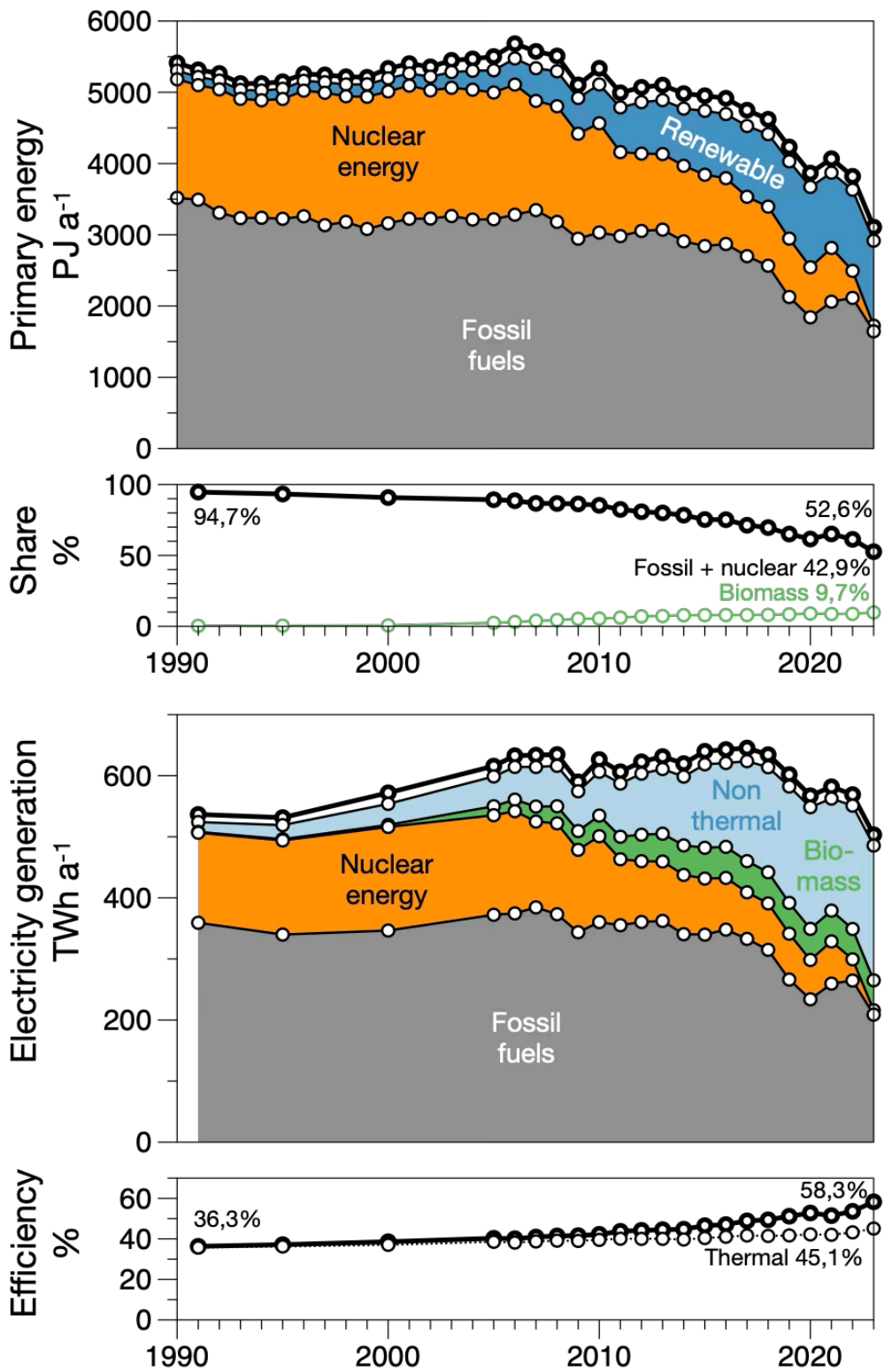

*FIGURE 4: ELECTRICITY GENERATION IN GERMANY*
*Top: Development of the use of primary energy (top) and the share of thermal power plants below. Below: Development of the resulting electricity generation and the overall efficiency, i.e. the ratio of electricity generation to primary energy used* (data sources: AGEB [1], BDEW [4]).

How and with what can we now achieve further increases in efficiency through modernisation, particularly in room heating and transport, where combustion still dominates?

At present, room heating - and hot water - is mainly generated by combustion, which represents 89 % of final energy consumption in households. By switching to heat pumps (see "The second law and energy conversions with heat"), the amount of energy used can be significantly reduced. Instead of generating heat by combustion at a very high temperature, which is not needed at all, the heat pump utilises electricity to maintain only a relatively small temperature difference. As the heat pump does work, it operates much more efficiently, but requires electricity. This means that one unit of electricity can generally be used to heat three to five times as much heat by a certain temperature difference (see "The second law and energy conversions with heat"). This distinguishes them from night storage heaters or infrared heaters, which generate heat more or less in a 1:1 ratio from electricity.

Currently, 89 % of 2275 PJ per year, i.e. 2025 PJ per year, are used for heat generation in households. By switching to heat pumps with an assumed coefficient of performance of 4 (see "The second law and energy conversions with heat"), the energy could be reduced to 506 PJ per year (Figure 5). Together with the 11 % consumption for other purposes, this results in an energy consumption of 756 PJ per year in households if combustion is replaced by electricity. Electricity demand would then increase by over 30 %, equivalent to 141 TWh per year.

In addition, the heat requirement can be further reduced if buildings are better insulated. The heat loss from buildings is directly related to the difference between the room temperature and the outside temperature. The better the walls and windows are insulated, the less room heating is required. For example, old buildings consume around ten times as much energy per square metre of living space for room heating as new buildings. Energy consumption can also be reduced by lowering the room temperature and using more intelligent controls, such as smart, programmable thermostats, so that not all rooms are heated at all times. If we assume that half of the living space can be reduced to a fifth through insulation and other measures, the energy demand in households is reduced to 452 PJ per year (Figure 5, right). The same considerations lead to similar reductions in the commercial, retail and service sectors.

In the mobility sector, electrification can also significantly reduce energy requirements. We have known a classic example of these conclusions for decades from rail transport. Basically, mobility means converting a form of energy into kinetic energy. In the case of the electric motor, electrical energy is used for this purpose with virtually no losses (although charging and discharging batteries is associated with certain losses). With the combustion engine, on the other hand, typically 80 % or more of the chemical energy contained in the fossil fuel is lost as waste heat, even if well-heated combustion engines can ideally achieve efficiencies of only 40 %. But ultimately it always comes down to the same thing - by avoiding waste heat when switching to e-mobility, the use of primary energy can be significantly reduced.

At present, 98% of the 2511 PJ per year of energy consumption in the transport sector is utilised through combustion. If this is reduced to 20% by switching completely to electromobility, energy use would fall to 542 PJ per year, but electricity demand would increase accordingly by around 30%, i.e. by 151 TWh per year.

The efficiency of mobility can also be significantly increased. In local transport, the energy requirement is characterised by acceleration work, i.e. associated with starting, braking, turning, accelerating, etc. The work to be done depends directly on the mass to be moved. The more mass is moved per person transported, the more work is required. For example, an off-road vehicle weighing 2.5 tonnes with one occupant requires significantly more energy per person to move than a fully occupied small car or a half-full tram. Electrification also allows some of the kinetic energy to be recovered when braking. Even less energy is needed when cycling, as the moving mass is mainly provided by the weight of the person riding the bicycle.

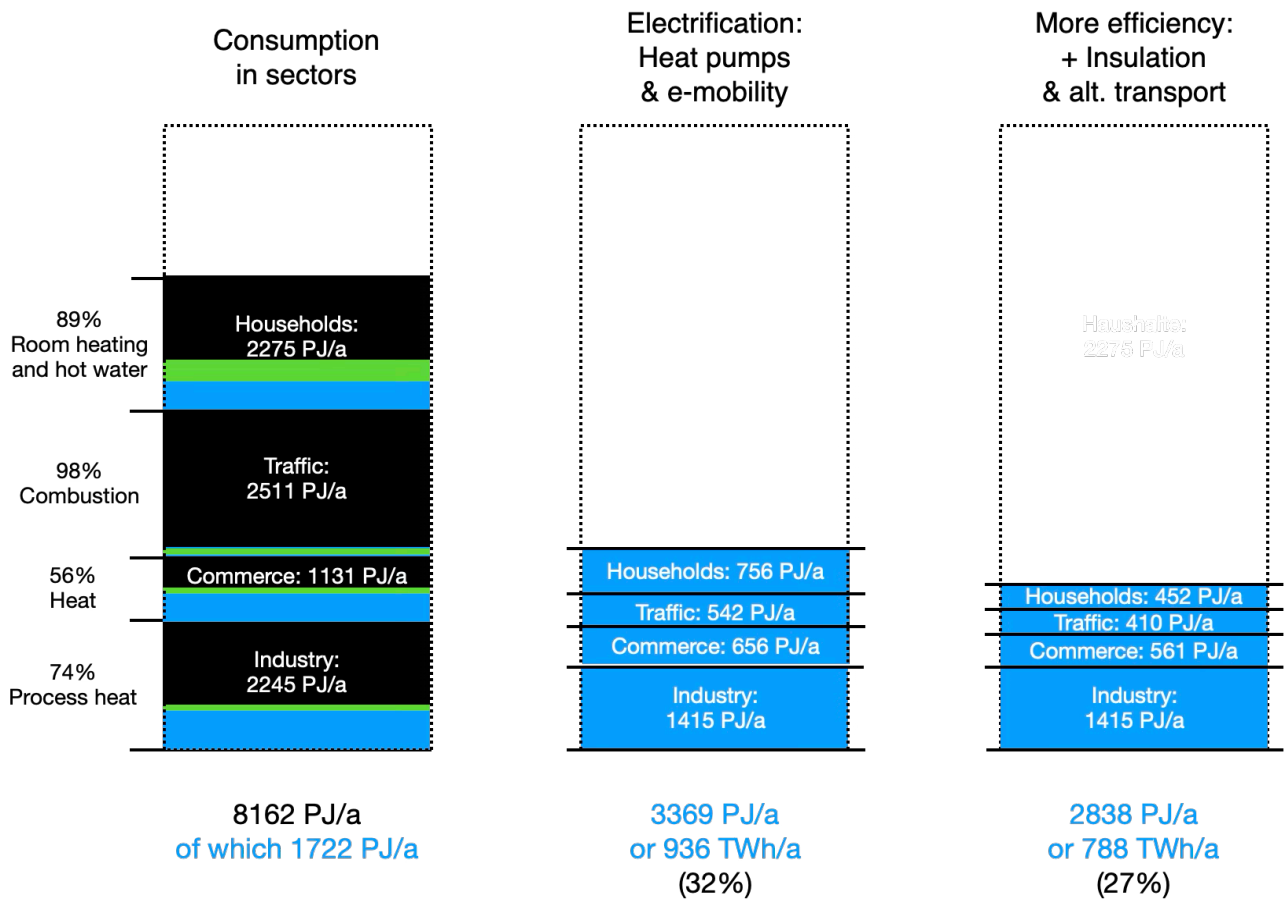

**FIGURE 5: PRIMARY ENERGY CONSUMPTION WITHOUT COMBUSTION**
*Current energy use in the sectors (left), the possible reduction with complete replacement by electricity-based technologies (centre) and through further measures such as insulation and non-use of cars for short journeys (right).*

Every year, Germans travel around $700 \cdot 10^9$ km, 85% of which is by car. 20 % of car journeys are for commuting - with around 50 % of work being less than 10 km away - and 40 % for shopping and leisure [5]. If cars were not used for half of these short journeys alone, this could reduce energy consumption by a further 24%, i.e. to 410 PJ per year. If mobility consisted of more public transport, cycling or walking, this would be significantly more efficient. More people could be moved with less use of primary energy.

In principle, energy can also be saved in industry through electrification. One example from paper production is the drying of paper using either heat from heat pumps or waste heat from existing thermal power plants instead of covering this heat requirement directly through combustion. For other processes, the switch to climate neutrality is more complex and not necessarily associated with efficiency improvements. For the sake of simplicity, we assume a 50% reduction in the use of process heat, which would reduce energy demand in the industrial sector to 1415 PJ per year.

*TABLE 1: A SCENARIO OF THE ENERGY TRANSITION*

|  | Capacity factor / % | Installed capacity / GW | Power generation / GW | Installed capacity 2024 |
|---|---|---|---|---|
| Photovoltaics | 12 | 350 | 42 | 90,3 |
| Wind onshore | 25 | 200 | 50 | 61,9 |
| Wind offshore | 35 | 70 | 25 | 8,9 |
| Total / GW |  |  | 117 |  |
| Total / TWh/a |  |  | 1021 |  |

Typical values for the expansion of photovoltaics and wind energy and how these could cover future energy consumption in the form of electricity, roughly estimated.

These estimates are of course rough and highly simplified - but above all they illustrate that primary energy demand can be reduced surprisingly significantly without lowering the standard of living. Energy use would follow the development that we described at the beginning for light and electricity generation: Our entire energy system would become more efficient. This electricity demand could then be met with an expansion of photovoltaics and wind on a scale (Table 1) that is used in current energy transition scenarios [6, 7].

## Conclusions

So we can summarise: The energy transition is above all a technological revolution that involves a fundamentally more efficient energy system in which combustion technology is replaced by electricity-based technologies. This can be explained directly with the second law of thermodynamics, including the associated maximum combustion temperature.

There is certainly no direct compulsion for greater efficiency, at least on the surface. But it is desirable in every respect:

1. Fundamental physical and technical: Recognising the basic laws of nature, such as the first and second laws of thermodynamics, is part of the ethos of the natural and technical sciences.

2. Economically, because less energy consumption always means lower costs. The potential savings are enormous: every year we spend around 80 billion euros on importing fossil fuels [7], which, as mentioned at the beginning, corresponds to more than 20 % of the federal budget. However, due to the low efficiency of combustion processes, a significant proportion - more than half - of this money is simply lost uselessly through chimneys, exhaust pipes and machine walls.

3. Less combustion also means much less air pollution and less global climate change. *The* British medical journal *The Lancet* has described climate change as the greatest global health risk in various large-scale publications [8].

4. Security is ensured by combining renewable energy sources with effective storage media such as batteries, thereby reducing imports of resources and reducing dependency on external supplies.

With so many benefits and sustainable effects of a successful transformation from combustion to electrification at all levels, there is really no longer any sensible reason not to do it.

## Summary

*The energy transition is not just about expanding renewable energy generation. It is also about switching to electricity-based technologies such as heat pumps and electromobility. Because they avoid combustion and heat as an intermediate step, they are much more efficient. This can be explained with the help of the maximum combustion temperature. Combustion could in principle produce energy with very low entropy. However, this temperature is not at all needed to generate room heating. For this reason, thermal power stations and combustion engines also lose a lot of energy via waste heat. By systematically switching to electricity-based technologies, these enormous waste heat losses can be avoided and Germany's primary energy consumption substantially reduced as a result. The energy transition is therefore not just about climate neutrality, but about a general modernisation of the energy system in Germany.*

## Keywords

Energy transition, thermodynamics, primary energy, consumption, combustion, efficiency, heat pump, electrification.

## About the authors

Axel Kleidon studied physics and meteorology at the University of Hamburg and Purdue University, Indiana, USA. After completing his doctorate at the Max Planck Institute for Meteorology, he conducted research at Stanford University in California and at the University of Maryland. Since 2006, he has headed the "Biospheric Theory and Modelling" group at the Max Planck Institute for Biogeochemistry in Jena. His research interests range from the thermodynamics of the Earth system to the natural limits of renewable energy sources.

Harald Lesch studied physics and philosophy at the Universities of Giessen and Bonn and gained his doctorate at the University of Bonn. He then carried out research in Heidelberg and Toronto. After his habilitation in 1994 at the University of Bonn, he became a professor of theoretical astrophysics at the LMU in 1995 and has been a lecturer in natural philosophy at the Munich School of Philosophy since 2002. He works on complex cosmic and terrestrial systems up to the natural limits of technological societies.